\documentclass[conference,compsoc]{IEEEtran}
\IEEEoverridecommandlockouts
\usepackage{cite}
\usepackage{amsmath,amssymb,amsfonts}
\usepackage{algorithmic}
\usepackage{graphicx}
\usepackage{textcomp}
\usepackage{xcolor,soul}

\def\BibTeX{{\rm B\kern-.05em{\sc i\kern-.025em b}\kern-.08em
    T\kern-.1667em\lower.7ex\hbox{E}\kern-.125emX}}
\begin{document}

\title{Comprehensive analysis of gene expression profiles to radiation exposure reveals molecular signatures of low-dose radiation response}

\author{
\IEEEauthorblockN{Xihaier Luo\IEEEauthorrefmark{1},
Sean McCorkle\IEEEauthorrefmark{1},
Gilchan Park\IEEEauthorrefmark{1},
Vanessa L\'opez-Marrero\IEEEauthorrefmark{1},
Shinjae Yoo\IEEEauthorrefmark{1},\\
Edward R. Dougherty\IEEEauthorrefmark{2},
Xiaoning Qian\IEEEauthorrefmark{1}\IEEEauthorrefmark{2},
Francis J. Alexander\IEEEauthorrefmark{1},
Byung-Jun Yoon\IEEEauthorrefmark{1}\IEEEauthorrefmark{2}
}
\IEEEauthorblockA{\IEEEauthorrefmark{1} Computational Science Initiative,
Brookhaven National Laboratory, Upton, NY
}
\IEEEauthorblockA{\IEEEauthorrefmark{2} Department of Electrical and Computer Engineering, Texas A\&M University, College Station, TX}
}

\maketitle

\begin{abstract}
There are various sources of ionizing radiation exposure, where medical exposure for radiation therapy or diagnosis is the most common human-made source. Understanding how gene expression is modulated after ionizing radiation exposure and investigating the presence of any dose-dependent gene expression patterns have broad implications for health risks from radiotherapy, medical radiation diagnostic procedures, as well as other environmental exposure. In this paper, we perform a comprehensive pathway-based analysis of gene expression profiles in response to low-dose radiation exposure, in order to examine the potential mechanism of gene regulation underlying such responses.
To accomplish this goal, we employ a statistical framework to determine whether a specific group of genes belonging to a known pathway display coordinated expression patterns that are modulated in a manner consistent with the radiation level.
Findings in our study suggest that there exist complex yet consistent signatures that reflect the molecular response to radiation exposure, which differ between low-dose and high-dose radiation.
\end{abstract}

\begin{IEEEkeywords}
Gene expression analysis, radiation biology, low-dose radiation response, pathway analysis.
\end{IEEEkeywords}

\section{Introduction}

Environmental threats constitute a major factor in determining a person's susceptibility to disease. With the progress of industrialization and modernization, radiation exposure has become one of the most serious environmental threats in today's world. Mounting evidence suggests that ionizing radiation is linked to the development of thyroid cancers, multiple myeloma, and myeloid leukemia in children and adults~\cite{national2006health}. It is well documented that the biological effects of ionizing radiation on mammalian cells are closely related to radiation doses and dose rates. In general, low-dose radiation exposure is far more common than high-dose radiation exposure because low-dose radiation can come from a variety of sources, including natural sources, cosmic rays, nuclear power, and various types of radioactive waste. However, in contrast to the more well-defined effects of high-dose radiation exposure, the biological effects and consequences of low-dose radiation and mixed exposures remain poorly understood~\cite{bonner2003low, mullenders2009assessing}.

Historically, the health risks associated with low-dose ionizing radiation exposure have been estimated by extrapolating from available high-dose radiation exposure data. However, the majority of the data come from experiments that used extremely high, even supra-lethal, doses. Extrapolating the results of such studies to physiologically relevant doses can thus be difficult~\cite{brenner2003cancer}. Furthermore, an increasing number of studies show that the biological reactions to high and low doses of radiation are qualitatively distinct, necessitating a direct examination of low-dose responses to better understand potential risks~\cite{tubiana2009linear}.

Genome-wide expression assays using microarrays or RNA sequencing can provide snapshots of transcriptional activities in a biological sample, hence studying the gene expression profiles under low doses of ionizing radiation can provide novel insights into the biological reactions to such radiation exposure. In fact, mining gene expression profiles has proven useful in understanding pathophysiological mechanisms, diagnosis and prognosis of complex diseases, and deciding on treatment plans. Several studies have demonstrated the effectiveness of using gene expression profiles for traditionally challenging problems, for instance, discriminating between different subtypes of a complex disease, such as cancer~\cite{golub1999molecular, alizadeh2000distinct}. Despite these successful applications, quantification and interpretation at the genetic level of the impact from radiation exposure on the risk of developing such diseases are still challenging. Especially, the small sample size of typical clinical data, on the other hand, frequently impedes meaningful analysis, making pattern discovery, disease marker identification, risk prediction, reproducibility, and validation extremely difficult~\cite{ideker2000testing, baldi2001bayesian}. Adjusting for multiple hypothesis testing is another critical issue for all microarray analysis methods. The similarities of such signatures across different sample types have not been demonstrated to be strong enough to conclude that they represent a universal biological mechanism shared by different sample types~\cite{braga2004cross, michiels2005prediction, ein2006thousands}. 

In recent years, scientists have gained a better understanding of the transcriptional response in cells to radiation exposure~\cite{khatri2012ten}. When cells are exposed to ionizing radiation, multiple signal transduction pathways are activated, making pathway activity a potentially powerful and informative approach for determining disease states. Furthermore, pathways, the most well-documented protein interactions, are known to closely reflect functional relationships related to molecular biological activities such as metabolic, signaling, protein interaction, and gene regulation processes. A growing body of research indicates that tasks such as class distinction based on differences in pathway activity can be more stable than distinction based solely on genes. For example,~\cite{lee2008inferring} incorporated pathway information into expression-based disease diagnosis and proposed a classification method based on pathway activities inferred for each patient. Later in~\cite{gatza2010pathway}, pathway activity patterns are used to describe a classification scheme for human breast cancer and to reveal complexity in intrinsic breast cancer subtypes. The probabilistic inference of differential pathway activity across different classes (e.g., disease states or phenotypes)
using probabilistic graphical models~\cite{Su2009} was shown to identify molecular signatures that can be used as robust and reproducible disease markers.
The marker identification method in~\cite{Su2009} was further extended in~\cite{khunlertgit2016incorporating}, where a novel algorithm for discovering robust and effective subnetwork markers in a human protein-protein interaction network that can accurately predict cancer prognosis and simultaneously discover multiple synergistic subnetwork markers.
It should be noted that at the heart of these pathway-based analyses is determining the activity of a given pathway based on the expression levels of the constituent genes.

The primary goal of this paper is to perform a comprehensive pathway-based analysis of gene expression profiles to investigate the differential time and dose effects, primarily in low-dose experiments, in order to uncover molecular signatures of low-dose radiation response. Towards this goal, we adopt the probabilistic pathway activity inference scheme in~\cite{Su2009}, where the pathway activity level is estimated from gene expression data via the use of a simple probabilistic graphical model. More specifically, the scheme estimates the log-likelihood ratio between different classes (e.g., different levels of radiation exposure) based on the expression level of each member gene. The log-likelihood ratios of the member genes in a given pathway are then aggregated for probabilistic inference of differential pathway activity. Through this analysis, we identify the most significantly differentially activated pathways in response to low-dose radiation. These pathways are investigated to determine the presence of consistent dose-dependent gene expression patterns. Our cross-validation experiments demonstrate that the proposed method can generate reliable and consistent pathway analysis results even with limited data.


\section{Data}

\subsection{Low-dose radiation gene expression data}

The goal of the current study is to identify potential molecular signatures underlying the biological response to low-dose ionizing radiation exposure through pathway-based analysis of gene expression profiles.
For this purpose, we conducted a thorough literature search and preliminary analysis to identify human gene expression data suitable for studying the low-dose radiation response. The gene expression dataset GSE43151\footnote{https://www.ncbi.nlm.nih.gov/geo/query/acc.cgi?acc=GSE43151} was identified to be the most suitable  for our study, in terms of sample size and the range of radiation levels that were considered.
Overall, GSE43151 contains gene expression measurements from 121 blood samples, where five healthy male donors provided 400 mL venous peripheral blood samples each~\cite{nosel2013characterization}.
A complete blood count was performed on each whole blood sample using an ADVIA Hematology System (Bayer HealthCare). The standard lymphocyte proportion of 16-45 percent was met by all samples. Heparin at a final concentration of 34 U ml$^{-1}$ was added to whole blood samples. The blood was then diluted 1:10 with Iscove's Modified Dulbecco's Medium (IMDM, Life Technologies). Finally, blood samples were incubated overnight at 37 Cina $5\%$ CO2 concentration.

For the \emph{ex vivo} irradiation, whole blood exposures were performed at the ICO-4000 facility (Fontenay-aux-Roses, France) with a Co source at a low dose rate (50 mGy min$^{-1}$). Exposures were carried out independently on each donor's blood sample. The kerma rate was calculated using a Physikalisch-Technische Werkst{\"a}tten (PTW) ionization chamber that was irradiated under the same conditions as the samples. Doses of 5, 10, 25, 50, 100, and 500 mGy were tested (See Table. \ref{tab:data}), as well as sham irradiated conditions. Following \emph{ex vivo} irradiation, blood samples were incubated at 37 degrees Celsius for 150, 300, 450, and 600 minutes in a $5\%$ CO2 atmosphere.

A density medium was used to collect CD4+ T lymphocytes for cell sorting. Following that, total RNA was extracted from CD4+ T lymphocytes using RNeasy Mini columns from the RNeasy Mini Kit (Qiagen) as directed by the manufacturer. For all RNA samples, the RIN (RNA integrity number) was calculated for assigning integrity values to RNA measurements. For gene expression assays, all RIN values were greater than the recommended value of 7.

Before performing the pathway analysis based on the GSE43151 gene expression dataset, all 121 samples in the dataset were normalized, filtered, and analyzed using GAGE in R software~\cite{luo2009gage}. Following the filtering step, a total of 10,875 probes were chosen, where the basic filtering criteria consisted of removing a probe when it was undetected in at least $75\%$ of the replicates considered.

\begin{table}[t!]
\centering
\begin{tabular}{cc}
\hline Dose Level & Number of Samples \\
\hline $0 \mathrm{~Gy}$ & 18 \\
$0.005 \mathrm{~Gy}$ & 16 \\
$0.01 \mathrm{~Gy}$ & 18 \\
$0.025 \mathrm{~Gy}$ & 18 \\
$0.05 \mathrm{~Gy}$ & 17 \\
$0.1 \mathrm{~Gy}$ & 18 \\
$0.5 \mathrm{~Gy}$ & 16 \\
\hline
\end{tabular}
\caption{Description of the gene expression dataset GSE43151 that was used to investigate the molecular signatures of low-dose radiation response in this study.}
\label{tab:data}
\end{table}

\begin{figure*}[t!]
    \centering
    \includegraphics[width=0.95\textwidth]{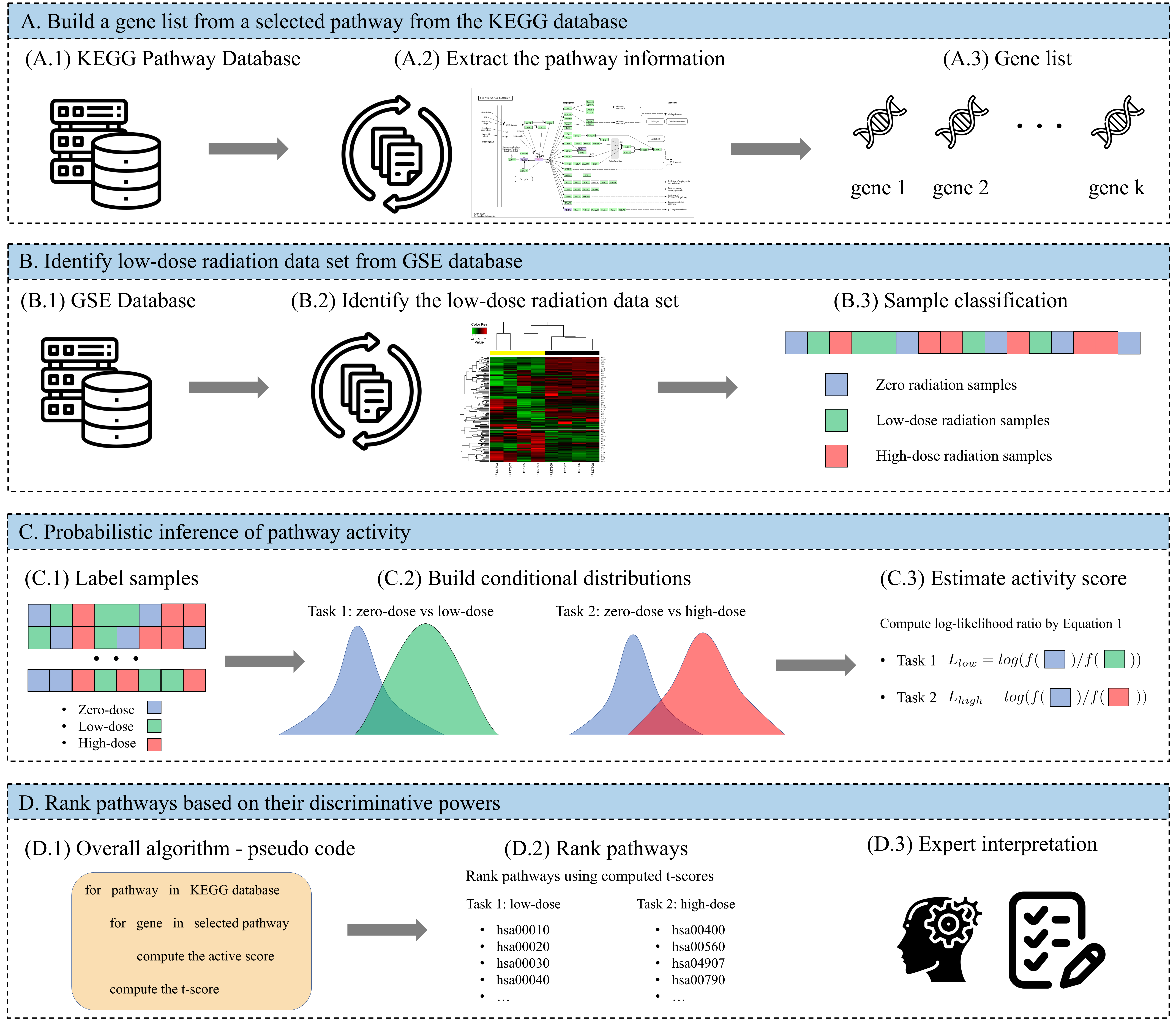}
    \caption{Overview of the pathway-based analysis of gene expression profiles in response to low-dose radiation exposure.}
    \label{fig:method}
\end{figure*}

\subsection{Pathway database}

We used the KEGG (Kyoto Encyclopedia of Genes and Genomes) database to obtain a reliable set of known biological pathways~\cite{kanehisa2000kegg}. KEGG is a collection of manually drawn pathway maps for understanding high-level functions and utilities of the biological system. The genomic information is maintained in the GENES database, which is a collection of gene catalogs for all fully sequenced genomes and some partially sequenced genomes with current annotations of gene functions. The PATHWAY database's higher-order functional information is augmented with a collection of ortholog group tables for information about conserved subpathways, which are frequently encoded by positionally related genes on the chromosome and are especially valuable in predicting gene functions. In our case, we identified 343 pathways relevant to the gene expression dataset GSE43151 from the available 548 KEGG pathway maps by discarding the pathways that did not contain any gene whose measurement was included in GSE43151.


\section{Methods}
\label{sec:methods}

In this section, we describe the technical details of the pathway-based gene expression data analysis procedure that was used to detect potential molecular signatures underlying low-dose radiation response. Figure~\ref{fig:method} provides an overview of the overall procedure.

\subsection{Pathway activity inference}
\label{sec:methods:activity}

To perform the pathway analysis, we first identified the genes whose  measurements were included in the gene expression dataset GSE43151 for the pathways of our interest. For every pathway, member genes that were missing in the given dataset were removed from the gene set. Consider a pathway $\mathcal{G}$ that consist of $n$ genes $\{g_k\}_{k=1}^{n}$ whose measurements were available in the dataset. In the context of binary classification, we assume that the expression level of gene $g_k$ ($k = 1, 2, \dots, n$) has a phenotype-dependent distribution. Let us denote the conditional probability density function (PDF) of gene $g_k$ expression level under phenotype 1 as $f^1_k(x)$ and the conditional PDF under phenotype 2 as $f^2_k(x)$ with $x$ representing the expression level of gene $g_k$. In our case, we classify radiation exposures into three categories: zero-dose, low-dose, and high-dose. We compare low-dose and high-dose samples separately to zero-dose samples, which means that if zero-dose samples are treated as phenotype~1, either low-dose or high-dose samples will be treated as phenotype~2.

After examining different probability distribution models, we assumed that both $f^1_k(x)$ and $f^2_k(x)$ are Guassian in this study.
Having these conditional PDFs, we can calculate the log-likelihood ratio (LLR) between the two phenotypes at a given expression level $x$ of gene $g_k$ as follows
\begin{equation}
    \label{eq: LLR}
    L_k(x) = \log [f_k^1(x)/f_k^2(x)]
\end{equation}
For any given gene $g_k$ in the pathway $\mathcal{G}$, the associated log-likelihood ratio $L_k(x)$ in \eqref{eq: LLR} indicates which phenotype is more likely based on the expression level $x$ of gene $g_k$. By combining the evidence--in the form of LLR--from all the member genes in the pathway, we can assess the overall  activity level of the pathway at hand to infer which of the two phenotypes the collective expression pattern of its member genes points to and how significantly so, as discussed in~\cite{Su2009}. More specifically, provided with a set $\{x_{j,k}\}_{j=1}^{m}$ of $m$ samples (i.e., gene expression measurements) for each gene $g_k$, we first calculated activity levels $\{S_{j}\}_{j=1}^{m}$ defined as 
\begin{equation}
    \label{eq: activity}
     S_j = \sum_{k=1}^{n} L_k(x_{j,k})
\end{equation} 
The activity level $S_j$ in \eqref{eq: activity} incorporates information from every gene in the pathway of interest and can be used to predict the phenotype (class label) based on the overall activation level of the given pathway in sample $j$.

Note that to calculate the log-likelihood ratio $L_k(x)$ in \eqref{eq: LLR}, we must first estimate the conditional PDF $f_k^{c}(x)$ for each phenotype $c \in \{ 1, 2\}$. We assume that the expression of gene $g_k$ under the phenotype $c$ follows a Gaussian distribution with a mean of $\mu_k^{c}$ and a standard deviation of $\sigma_k^{c}$. These parameters were calculated using all of the available samples that correspond to the phenotype $c$. After that, the estimated conditional PDFs can be utilized to compute the log-likelihood ratios. In practice, we often have insufficient training data to estimate the PDFs of $f_k^{1}(x)$ and $f_k^{2}(x)$ with confidence. As a result, the computation of the log-likelihood ratio may be sensitive to relatively small changes in the gene expression levels. 
To alleviate this issue, we normalized the data as recommended in~\cite{Su2009}.
Namely, $L_k(x)$ was normalized to obtain $\widehat{L}_k(x)$ as follows
\begin{equation}
    \label{eq: normalize}
    \widehat{L}_k(x) = \frac{{L}_k(x) - \mathbb{E}[{L}_k(x)]}{\sqrt{\mathbb{E} [ ({L}_k(x) - \mathbb{E}[{L}_k(x)])^2 ] }} \, .
\end{equation}
While the use of \eqref{eq: LLR} and \eqref{eq: activity} without normalization for inferring the pathway activity level would be equivalent to using a Naive Bayes model (NBM) for classifying the phenotype (class label) given the expression profile of the member genes that belong to a given pathway, this normalization step in \eqref{eq: normalize} makes the pathway activity scoring scheme diverge from the traditional NBM.

\subsection{Pathways as potential markers for discriminating low-dose response from high-dose response}
\label{sec:methods:marker}

To examine the ability of a pathway to discriminate between two phenotypes, we computed the $t$-test statistics scores using the activity levels $S_j$ for all member genes (as defined in \eqref{eq: activity}) and averaged the absolute value of the $t$-test scores to compute an aggregated differential activity score. The aggregated score--which we refer to as the \textit{pathway activity score}--was then used as an indicator of the pathway's discriminative power~\cite{tian2005discovering}. It should be noted that low-dose and high-dose samples were analyzed separately to detect most strongly differentially activated pathways under each radiation exposure level. We had three types of samples: zero radiation, low-dose radiation (0.005 Gy to 0.1 Gy), and high-dose radiation (0.5 Gy). Despite the fact that different low-dose levels of ionizing radiation have been tested, we treated all dose levels between 0.005 Gy and 0.1 Gy as the same type (i.e., low-dose radiation). Based on this categorization, we ranked all relevant KEGG pathways to based on the strongest differential pathway activity between zero-dose against low-dose radiations, and separately, based on zero-dose against high-dose radiations. This is illustrated in Fig.~\ref{fig:method}(C).

\section{Results}
\label{sec:results}

\subsection{Pathway analysis results}
To begin, we evaluated all relevant pathways in the KEGG database and ranked the pathways based on their discriminative power following the procedures elaborated in Sec.~\ref{sec:methods} and illustrated in Fig.~\ref{fig:method}. In particular, we ranked the pathways based on their discriminative power, assessed based on the aggregated differential activity score obtained by averaging the absolute value of the $t$-test scores of the member genes in a given pathway~\cite{tian2005discovering} and estimating the $p$-value.

\begin{figure}[t!]
    \centering
    \includegraphics[width=0.485\textwidth]{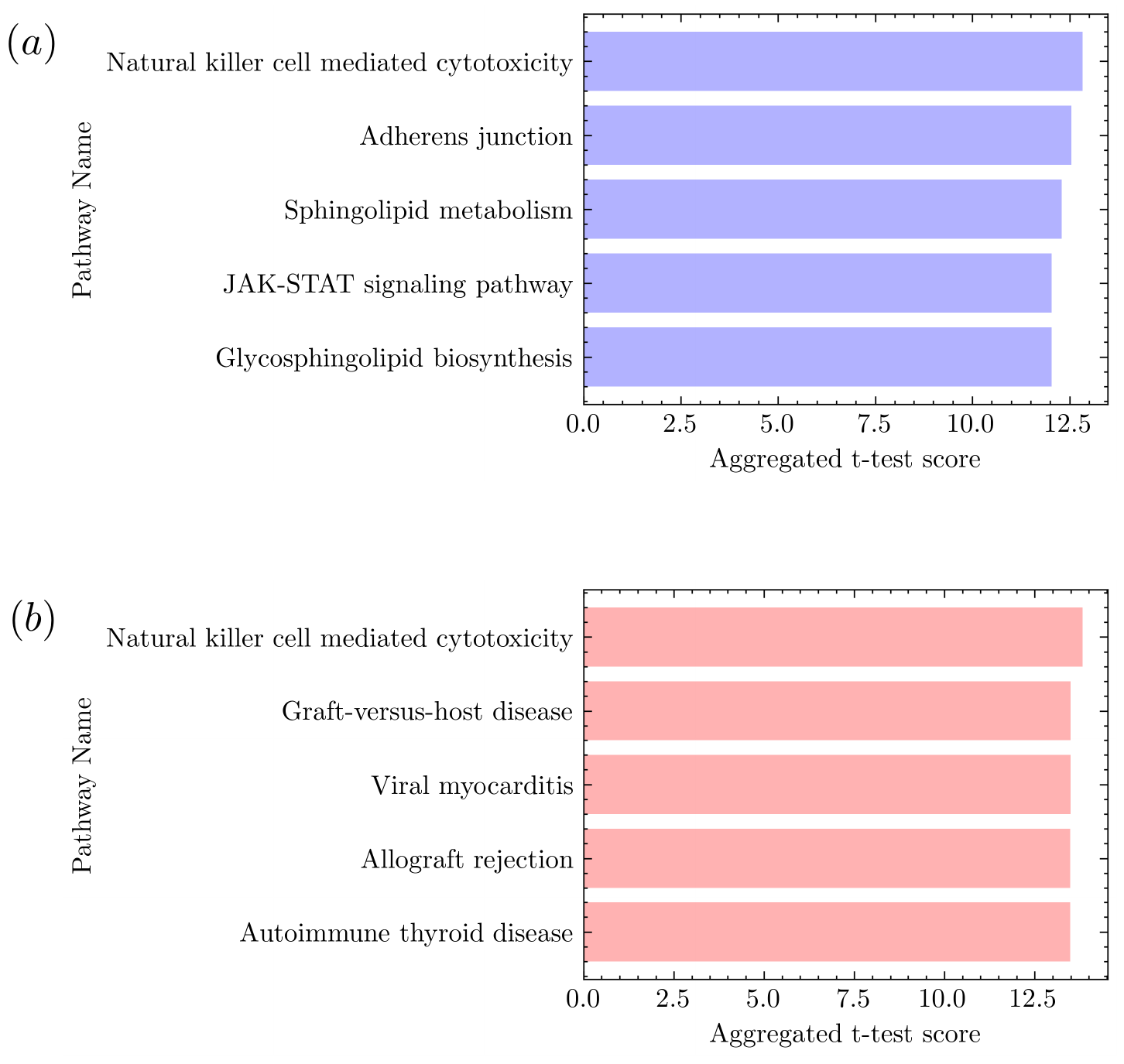}
    \caption{Ranking of most differentially activated pathways and their discriminative power in terms of the pathway activity score.  (a) Top differentially activated pathways under low-dose radiation exposure. The aggregated $t$-test scores reflect the discriminative power of the pathways for discriminating between zero-dose and low-dose samples. (b) Top differentially activated pathways for high-dose radiation exposure (zero-dose vs high-dose). Comparison between (a) and (b) show a significant difference between the list of top pathways that are differentially activated under low-dose radiation and those under high-dose radiation.}
    \label{fig:ranked_pathway}
\end{figure}

Fig.~\ref{fig:ranked_pathway}(a) shows the top five pathways that have been identified as being the most deferentially activated in the presence of low-dose radiation.

The top pathway was associated with \textit{Natural killer cell mediated cytotoxicity}, focusing on natural killer cells, which are innate immune system lymphocytes involved in early defenses against both allogeneic and autologous cells. Many studies have been conducted to investigate the direct effects of low-dose ionizing radiation (LDIR) on natural killer cells and the potential mechanism~\cite{yang2014low, jeong2018effect}. The results of the experiments showed that a simplified strategy based on LDIR leads to effective expansion and increased activity of natural killer cells, providing a novel approach for adoptive cellular immunotherapy.

The second pathway is related to \textit{Adherens junction (AJ)}, which is the most common type of intercellular adhesion. AJ initiates and maintains cell adhesion while also controlling the actin cytoskeleton. In~\cite{sandor2014low}, three types of junctional proteins were chosen for immunohistochemical labeling, and experimental results showed that not only high, but also low and moderate doses of cranial irradiation increase cerebral vessel permeability in mice. \emph{In-vitro} studies showed that irradiation alters junctional morphology, reduces cell number, and causes senescence in brain endothelial cells. Another study~\cite{shukla2016rapid} discovered that gamma-radiation, even at low doses, rapidly disrupts tight junctions, adherens junctions, and the actin cytoskeleton, resulting in barrier dysfunction in the mouse colon \emph{in vivo}. Radiation-induced epithelial junction disruption and barrier dysfunction are mediated by oxidative stress, which can be mitigated by NAC supplementation prior to IR.

Another pathway linked to \textit{Sphingolipid metabolism} was also highly ranked. Sphingolipids, a type of membrane lipid, are bioactive molecules that play a variety of roles in fundamental cellular processes such as cell division, differentiation, and cell death. Many studies on the effect of sphingolipids on cancer treatment have been conducted. Microbeam radiation can induce radiosensitivity in elements within the cytoplasm, according to~\cite{shao2004targeted}. The effect could be inhibited by agents that disrupt the formation of lipid rafts (filipin), demonstrating once again that membranes could be a target of ionizing radiation. The authors of~\cite{truman2014evolving} concluded that, while other pathways are activated to induce radiation or chemoresistance, sphingolipids play a significant role.

The JAK-STAT signaling pathway and Glycosphingolipid biosynthesis have also been revealed to be very important in the study of radiation effects. For example, erythropoietin (EPO), which was originally identified as an erythrocyte growth factor, is now used to treat anemia and fatigue in cancer patients receiving radiation therapy and chemotherapy. The study in~\cite{lai2005erythropoietin} demonstrated previously unknown EPO-mediated HNSCC cell invasion via the Janus kinase (JAK)-signal transducer and activator of transcription (STAT) signaling pathway. On the other hand, the findings in ~\cite{aureli2014exploring} suggest that glycosphingolipid biosynthesis on the cell surface contributed to the activation of ionizing radiation-induced apoptosis via ceramide production. The functional importance of this pathway to eradicating cancer cells with ionizing radiation has been proven, with sphingolipid breakdown activated as a mechanism of ceramide formation after cell irradiation.

In a similar manner, Fig.~\ref{fig:ranked_pathway}(b) shows the top five pathways that have been identified as being most differentially activated in the presence of high-dose radiation. The genes found in the identified pathways are closely related to the radiotherapy regimen. Graft-versus-host disease (GVHD), for example, is a fatal complication of allogeneic hematopoietic stem cell transplantation in which immunocompetent donor T cells attack genetically diverse host cells. Many clinical studies have found a link between GVHD severity and radiation dose, with more severe GVHD after conditioning regimens that included radiation therapy compared to those that only included chemotherapy~\cite{hill2000primacy, blazar2012advances}. Another example is allograft rejection. By definition, the recipient's alloimmune response to nonself antigens expressed by donor tissues causes allograft rejection. According to research, the complex pathophysiology involves host tissue damage caused by the conditioning regimen (chemotherapy and/or irradiation)~\cite{van2002cytolytic}. After nonmyeloablative conditioning with low-dose irradiation, the use of recombinant fusion protein promotes mixed lymphoid chimerism.

Interestingly, we can see that there is  relatively small overlap between the set of pathways there were most responsive to low-dose radiation exposure and those that were responsive to high-dose radiation exposure. For example, as shown in Fig.~\ref{fig:ranked_pathway}, only one pathway (i.e., \textit{Natural killer cell mediated cytotoxicity}) was among the top 5 differentially activate pathways under both low-dose and high-dose radiation. However, we can see more pathways in common as we go down the list further. For example, when we compare the top ten pathways that are the most responsive to low-dose and high-dose radiation exposure, we find four common pathways: \textit{Natural killer cell mediated cytotoxicity}, \textit{Adherens junction}, \textit{Glycosphingolipid biosynthesis}, and \textit{Antigen processing and presentation}.

\begin{figure*}[t!]
    \centering
    \includegraphics[width=0.95\textwidth]{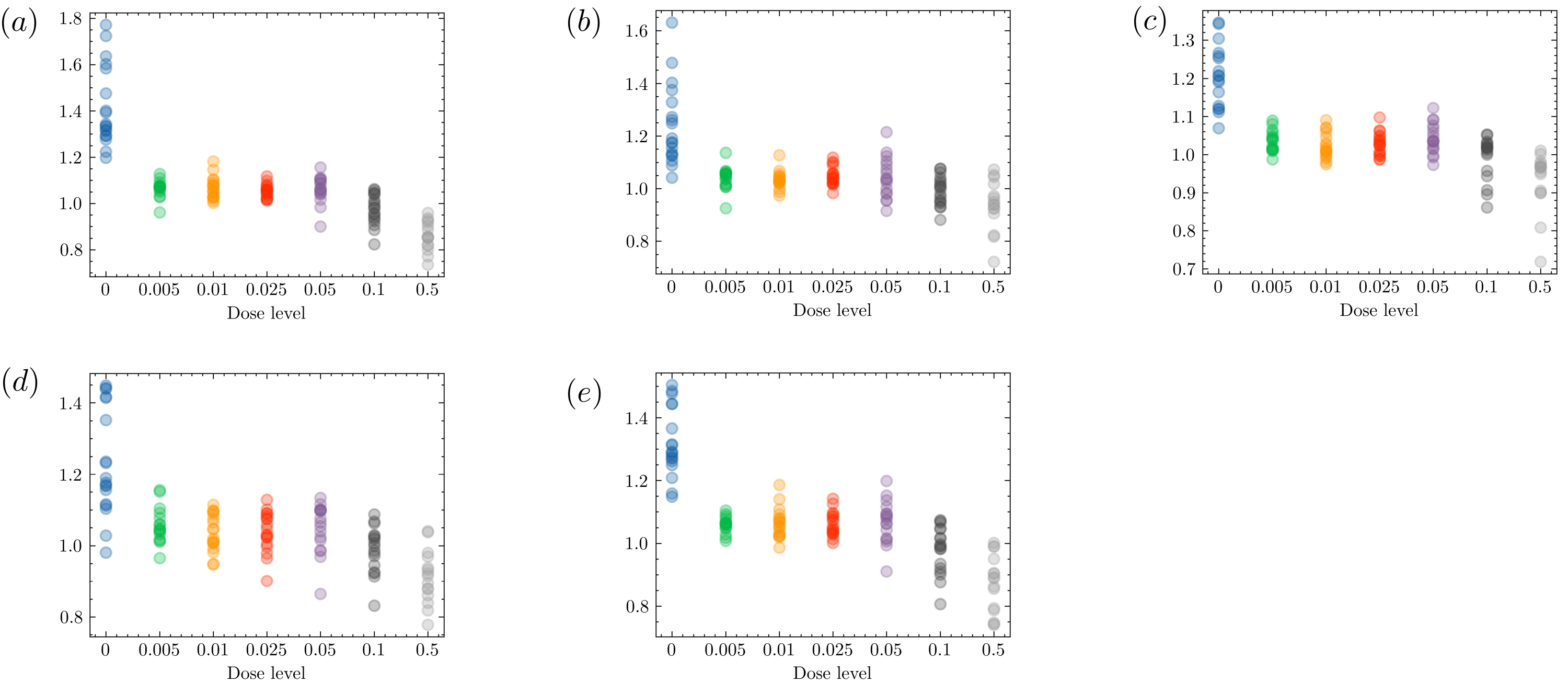}
    \caption{The pathway activity level measured in terms of the aggregated log-likelihood ratios (LLRs) in response to different levels of radiation exposure. Dose-dependent activity level is shown for the top five pathways that were most differentially activated under low-dose radiation exposure. (a) Natural killer cell mediated cytotoxicity (b) Adherens junction (c) Sphingolipid metabolism (d) JAK-STAT signaling pathway (e) Glycosphingolipid biosynthesis. All plots in (a)--(e) for the top low-dose response pathways display similar trends, where the differential activity levels reflecting the presence of potential molecular signatures of low-dose radiation response decrease as the radiation dose level increases.}
    \label{fig:GSE43151_low}
\end{figure*}

\begin{figure*}[t!]
    \centering
    \includegraphics[width=0.95\textwidth]{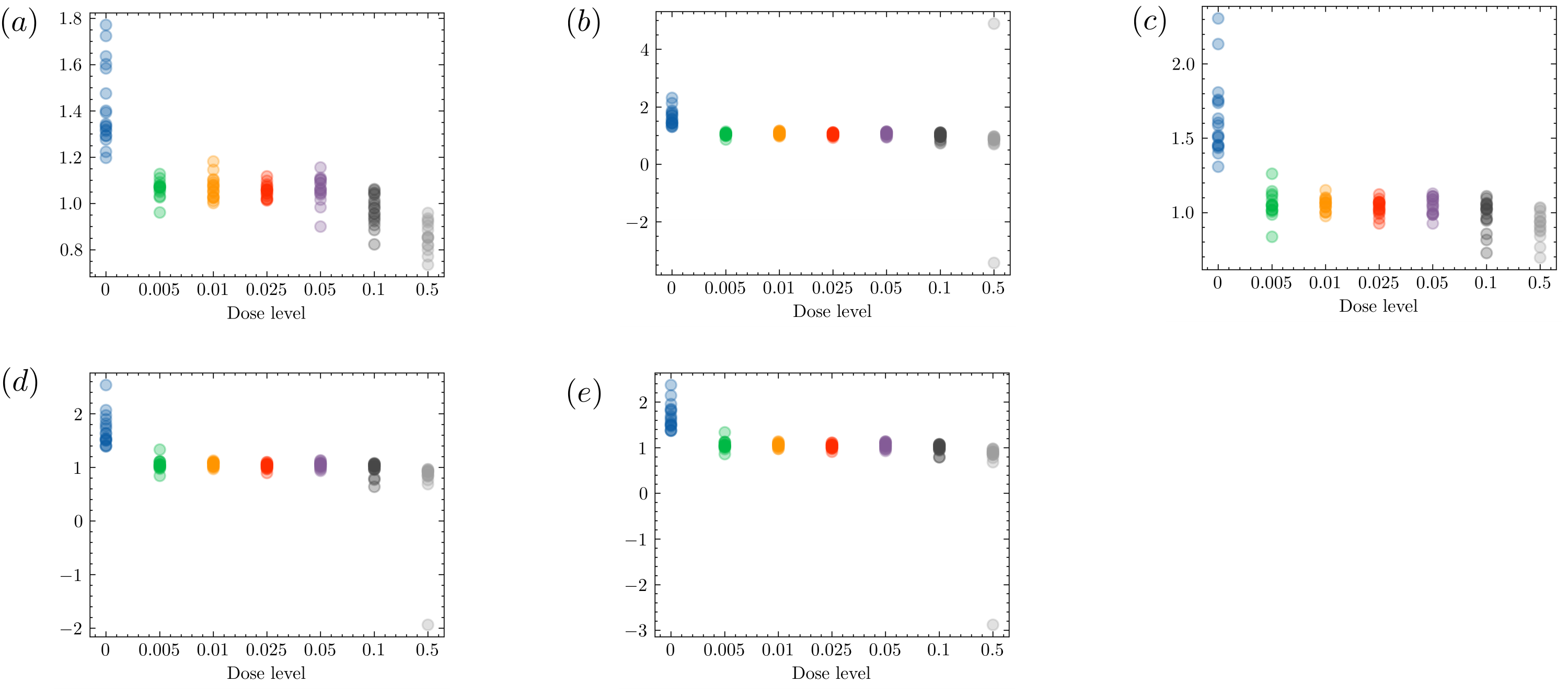}
    \caption{The pathway activity level measured in terms of the aggregated log-likelihood ratios (LLRs) in response to different levels of radiation exposure. As before, dose-dependent activity level is shown for the top five pathways that were most differentially activated under high-dose radiation exposure. (a) Natural killer cell mediated cytotoxicity (b) Graft-versus-host disease (c) Viral myocarditis (d) Allograft rejection (e) Autoimmune thyroid disease. Except for the top pathway in (a), the differential activity levels reflecting the presence of potential molecular signatures of high-dose radiation response do not significantly change as the radiation dose level increases. This implies that the pathways that are responsive to high-dose radiation exposure may not be substantially perturbed under relatively lower-dose radiation exposure.}
    \label{fig:GSE43151_high}
\end{figure*}

\subsection{Differential dose effect on radiation responsive pathways}


Next, we investigated the differential dose effects on the top pathways that were most responsive to either low-dose or high-dose radiation exposure.
As noted earlier in Sec.~\ref{sec:methods:activity}, the probabilistic pathway activity inference scheme~\cite{Su2009}, which we adopted in this current study, is equivalent to using a simple probabilistic graphical model (PGM)--namely, a NBM--when we use \eqref{eq: activity} for calculating the pathway activity score based on the LLRs of the member genes belonging to the pathway.
We wanted to find out whether this PGM constructed to detect the presence of low-dose (or high-dose) radiation exposure yields consistent activity inference results as the radiation dose level changes.

Figure~\ref{fig:GSE43151_low} shows the inference result based on the PGM trained to discriminate between \textit{zero-dose} and \textit{low-dose} samples. The $y$-axis shows the aggregated LLRs and the $x$-axis corresponds to the radiation dose level.
For each dose level, the dots show the distribution of the pathway activity scores for all samples radiated at the given dose level. The results are shown for the top five pathways that were found to be most responsive to low-dose radiation.
As we can see in Fig.~\ref{fig:GSE43151_low}, all low-dose responsive pathways yielded similar trends, where the inferred differential activity levels generally decreased as the radiation exposure level increased. These results imply that these pathways, and the gene expression profiles of the members therein, may reflect potential molecular signatures underlying the biological response to low-dose radiation exposure.

We carried out a similar analysis based on the top five high-dose radiation response pathways that were identified in our study. The analysis results are summarized in Fig.~\ref{fig:GSE43151_high}.
As before, the $y$-axis shows the pathway activity score obtained by aggregating the LLRs of the member genes in the pathway at hand. It should however be noted that, in this case, the LLR is obtained by comparing the likelihood ratios between zero-dose response and high-dose response. The resulting PGM is therefore trained to discriminate between \textit{zero-dose} samples and \textit{high-dose} samples. 
Interestingly, except for the first pathway (i.e., \textit{Natural killer cell mediated cytotoxicity}), which was the top-ranked pathway in both low-dose as well as high-dose differential activity analysis (see Fig.~\ref{fig:ranked_pathway}), the pathway activity levels did not change significantly as the dose level increased. Considering that the pathway activity scores reflect the presence of potential molecular signatures of high-dose radiation response, this may imply that these top pathways that were responsive to high-dose radiation exposure might not be substantially perturbed when the radiation dose level is relatively low.


\subsection{Reproducibility of the identified pathways}

\begin{figure}[t!]
    \centering
    \includegraphics[width=0.49\textwidth]{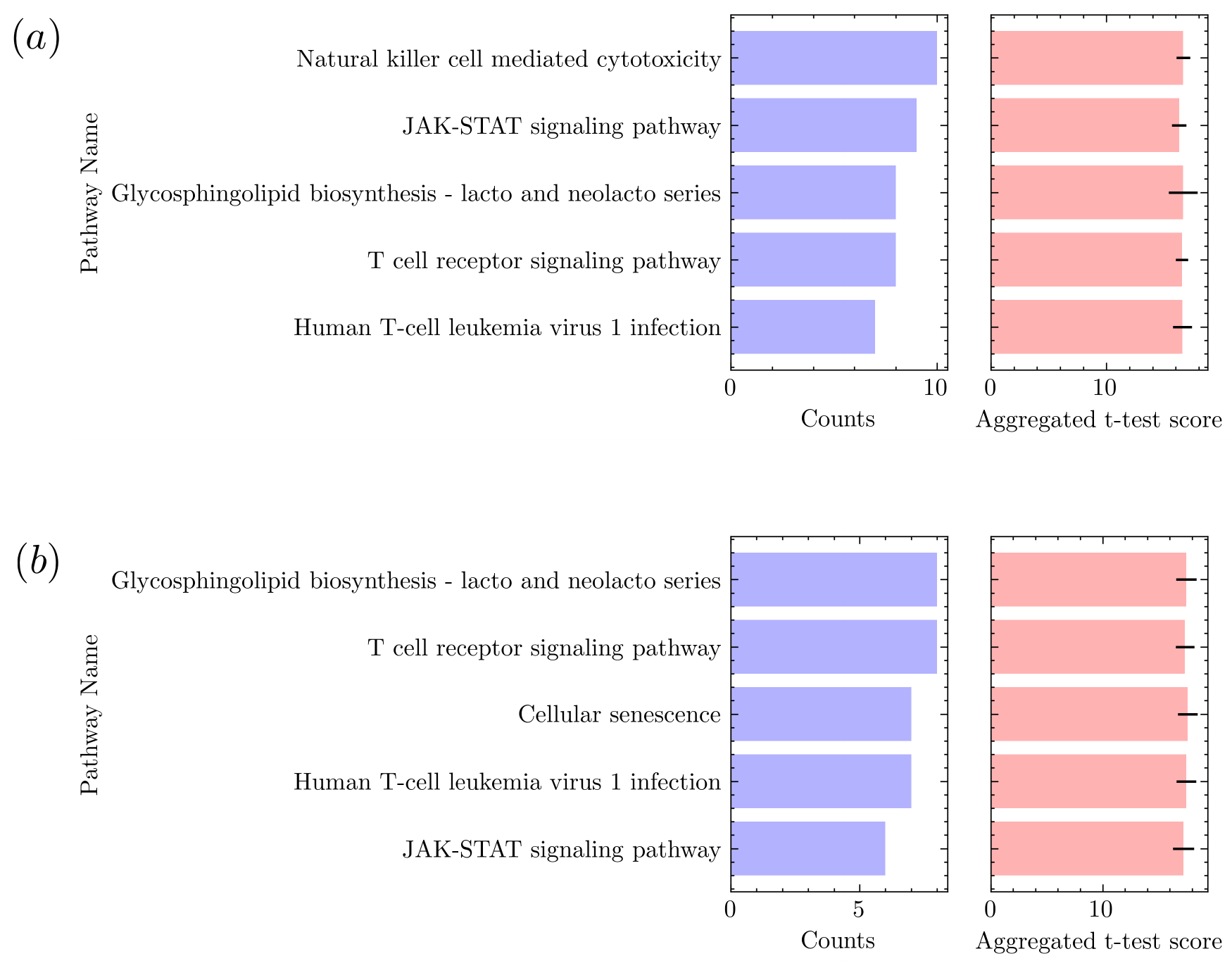}
    \caption{Cross validation results of the top ranked pathways. (a) Cross-validation results for pathways most responsive to low-dose radiation. (b) Cross-validation results for pathways most responsive to high-dose radiation.}
    \label{fig:cross_validation}
\end{figure}


We conducted cross-validation experiments to assess the reproducibility of pathway analysis results and the significance of the identified pathways. To begin the experiment, we randomly selected $70\%$ of zero-dose, low-dose, and high-dose samples, and we repeated this process ten times, taking into account the total size of our dataset. The top-ranked pathways identified by the algorithm are depicted in Fig. \ref{fig:cross_validation}. Because the different sample selection introduces randomness, we first counted the show-up cases of pathways from the top ten most activated pathways. Then, we ranked our cross-validation results based on the total number of counts (shown in blue color). We also computed the mean and standard deviation of the aggregated $t$-test scores for each pathway (shown in red color). The cross-validation experiments for low-dose radiation responsive pathways are shown in see Fig.~\ref{fig:cross_validation}(a). As we can see, Fig.~\ref{fig:cross_validation}(a) demonstrates the consistency of the identified pathways when compared to the results originally obtained using the whole dataset (see Fig.~\ref{fig:ranked_pathway} for comparison). Pathways \textit{Natural killer cell mediated cytotoxicity} and \textit{JAK-STAT signaling pathway}, for example, have been identified as being highly related to low-dose radiation response. We suspect that the difference is due to the radiation dose level. As previously discussed, we discovered a direct relationship between dose level and activation. Such differences are expected in a mixed and random combination of different dose levels.

Noticeably, such consistency was not observed in the high-dose experiments shown in Fig.~\ref{fig:cross_validation}(b). In many top-ranked pathways, as shown in Fig.~\ref{fig:GSE43151_high}, there is a weak distinction between high-dose samples. The last column, which represents the distribution of the calculated aggregated t-test scores of high-dose samples, in particular, shows a narrow-band distribution (See Fig.~\ref{fig:GSE43151_high}(b), (d), and (e)). Because the calculated statistical scores are so close, when randomness is introduced into data sampling, the cross-validation results in Fig.~\ref{fig:cross_validation}(b) appear more random. To validate this, we expanded our ranked pathway list to the top 30 pathways and found a larger number of overlapping pathways between the experiments using full dataset and the cross-validation experiments using only $70\%$ of the dataset. In this case, the average ranking of the pathways \textit{Natural killer cell mediated cytotoxicity} and \textit{Allograft rejection}, for example, were 17th and 22nd, respectively. It should be noted that the radiation dose level that we categorized as ``high-dose'' in this study is still relatively low. We expect that gene expression analysis of samples that underwent higher-dose radiation exposure may result in more consistent pathway identification results with clear molecular signatures.    

\begin{figure}[h]
    \centering
    \includegraphics[width=0.47\textwidth]{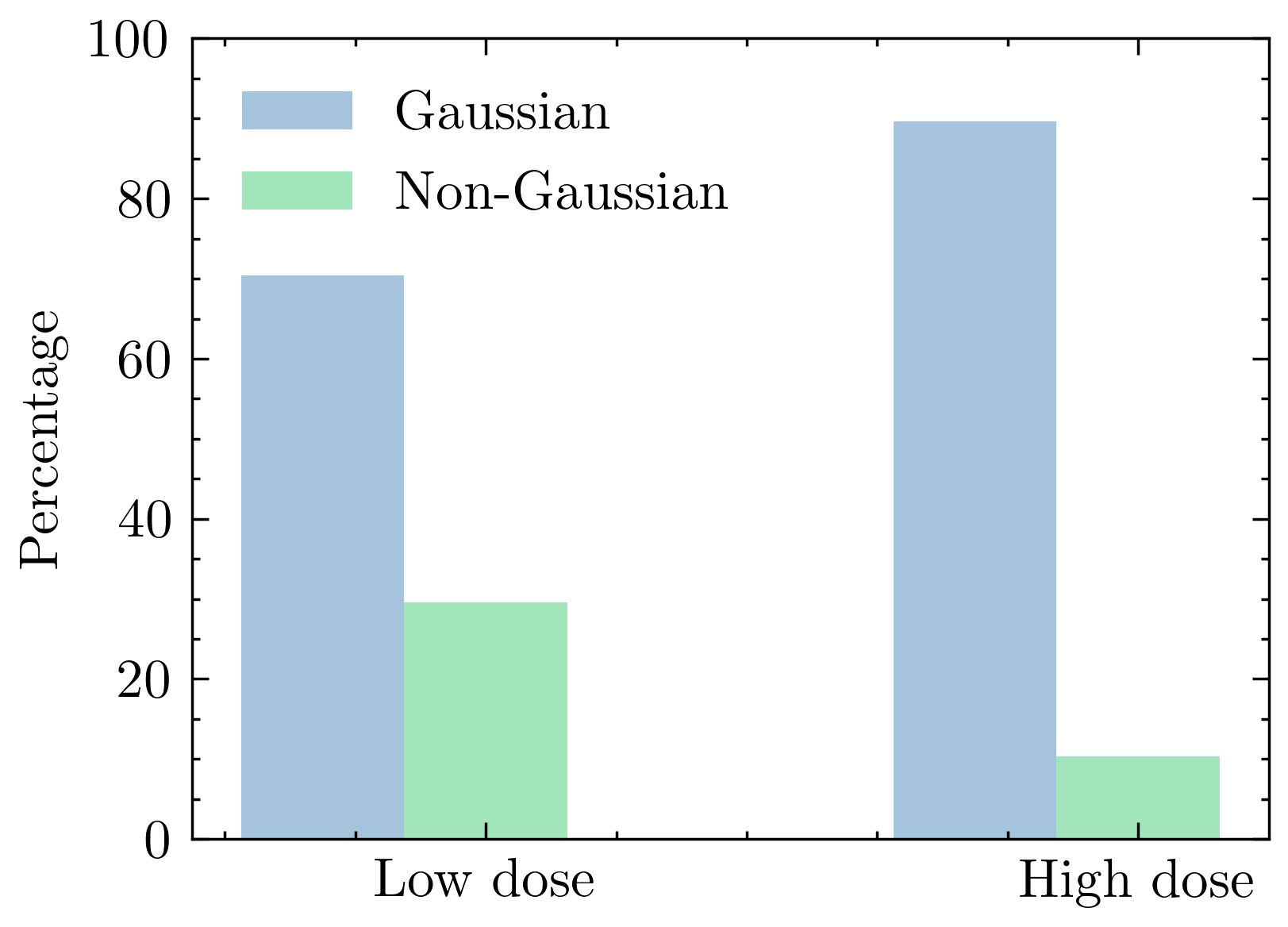}
    \caption{Kolmogorov-Smirnov (KS) test results.  We checked the normality of the gene expression values in low-dose and high-dose samples using the KS test. Results indicate that the Gaussian assumption holds in most cases.}
    \label{fig:Kolmogorov_Smirnov_test}
\end{figure}

Finally, we also investigated the assumption regarding the conditional distribution of the gene expression values. We used the one-sample Kolmogorov-Smirnov (KS) test to determine the goodness of fit. The test compares the underlying distribution $F(x)$ of a sample to a given distribution $G(x)$, which in our case is a Gaussian distribution. The null hypothesis holds that the two distributions are identical, with $F(x)=G(x)$ for all x; the alternative holds that they are not. We classify the samples as having a Gaussian distribution if the P-value is greater than 0.05; otherwise, they have a non-Gaussian distribution. Figure~\ref{fig:Kolmogorov_Smirnov_test} depicts the computed results, which show that 70.45 percent of the low-dose samples and 89.63 percent of the high-dose samples adhere to the Gaussian assumption. This indicates that during the pathway analysis, it is appropriate to assume that the conditional distribution of the gene expression data is Gaussian.

\section{Conclusion}
The current study aimed to unveil molecular signatures of biological responses exposed to low or very low doses of ionizing radiation through pathway-based analysis of  genome-wide expression profiles.
Gene expression patterns under the radiation exposure at six different dose levels ranging from 5 mGy to 500 mGy were investigated, where the measurements in the original study~\cite{nosel2013characterization} were made using blood samples obtained from five different donors during five independent irradiation sessions. Our investigation was conducted at the pathway level, as pathway-based gene expression analysis is known to yield more robust and reproducible results and as it may shed light on potential molecular mechanisms underlying low-dose radiation response. To determine the differential activity level of a given pathway under different levels of radiation exposure, a probabilistic pathway activity inference scheme was adopted that aggregates the log-likelihood ratios (LLRs) of the member genes in a given pathway to infer its differential activity. This allows robust detection of pathways, whose member genes display possibly subtle yet consistent coordinated expression patterns in response to low-dose radiation exposure. We searched through the KEGG database to prioritize pathways based on their differential activity levels modulated by low-dose radiation exposure. Our analysis identified the top pathways that may be associated with low-dose radiation response. Findings in this study reflect the complicated nature of the biological response to low-dose ionizing radiation, as well as the fact that low-dose exposures affect many different gene pathways that are not significantly altered after higher doses of radiotherapy.

One limitation of the current study is the small sample size of the analyzed dataset (GSE43151). While it has been challenging to find large-scale human gene expression data under low-dose radiation exposure, should such data be available in the future, their analysis would shed further light onto the unique molecular signatures of low-dose radiation response. Furthermore, the pathway activity level inference scheme in \eqref{eq: activity} makes specific modeling assumptions, upon which the derived results depend. In fact, the adopted scheme~\cite{Su2009} assumes that the gene expression levels of the member genes in a given pathway are conditionally independent given the class label (e.g., presence/absence of radiation exposure as was considered in the current study) and follow Gaussian distributions. Although we carried out some preliminary validation of this modeling assumption (e.g., see Fig.~\ref{fig:Kolmogorov_Smirnov_test}), it would be also worth validating the pathway analysis results using other methods~\cite{Han2015esea, Rydenfelt2020speed2}, which may be potentially pursued in our future studies.

\section*{Acknowledgements}

This work is supported by the U.S. Department of Energy, Office of Science, RadBio program under Award KP1601011/FWP CC121.

\bibliographystyle{IEEEtran}
\bibliography{ref}

\end{document}